\documentstyle[12pt,aasms4]{article}

\slugcomment{submitted to {\em The Astrophysical Journal}}

\lefthead{Tucker et al.}
\righthead{1E0657-56: A CONTENDER FOR THE HOTTEST KNOWN CLUSTER OF GALAXIES}

\begin{document}

\title{1E0657-56: A CONTENDER FOR THE HOTTEST KNOWN CLUSTER OF GALAXIES}

\author{W. Tucker\altaffilmark{1}, P. Blanco, S. Rappoport}
\affil{University of California, San Diego}
\affil{Center for Astrophysics and Space Sciences}
\affil{9500 Gilman Drive, Code 0111}
\affil{La Jolla, CA 92093}

\author{L. David, D. Fabricant, E.E. Falco, W. Forman}
\affil{Harvard-Smithsonian Center for Astrophysics}
\affil{60 Garden Street}
\affil{Cambridge, MA 02138}

\author{A. Dressler}
\affil{Carnegie Observatories}
\affil{813 Santa Barbara Street}
\affil{Pasadena, CA 91101}

\author{M. Ramella}
\affil{Osservatorio Astronomico di Trieste}
\affil{via G.B. Tiepolo, 11,I-34131}
\affil{Trieste, Italy}


\altaffiltext{1}{Also at the Harvard-Smithsonian Center for Astrophysics}


\begin{abstract}
We identify the extended Einstein IPC X-ray source, 1E0657-56, with a
previously unknown cluster of galaxies at a redshift of
$z=0.296$. Optical CCD images show the presence of a gravitational arc 
in this cluster and galaxy spectra yield a cluster velocity dispersion of
$1213^{+352}_{-191}$~km~s$^{-1}$.  X-ray data obtained with the ROSAT HRI and
ASCA indicate that 1E0657-56 is a highly luminous cluster in which a
merger of subclusters may be occurring.  The temperature of the hot
gas in 1E0657-56 is $\rm{kT}=17.4 \pm ~2.5 \rm{keV}$, which makes it
an unusually hot cluster, with important cosmological implications.
\end{abstract}




%

\section{INTRODUCTION}

Measurements of the gas temperature in clusters of galaxies is a
potentially powerful tool for testing cosmological models since the
gas temperature provides a direct measurement of the cluster mass.
The cluster mass function is directly traceable to the underlying
cosmology and can be used to determine the normalization and slope of the initial
mass fluctuation spectrum in the early Universe (Peebles et al. 1989,
Henry \& Arnaud 1991, Lilje 1992, Bartlett \& Silk 1993, White et
al. 1993, Colafrancesco \& Vittorio 1994, Hattori et al. 1995, Eke et
al. 1995, Eke et al. 1996, Pen 1996, Viana \& Liddle 1996, Oukbir et
al. 1997, Oukbir \& Blanchard 1997, Bahcall et al. 1997).  The
usefulness of this method is limited by the uncertainty in the
parameters used to fit the theory of cluster formation to the observed
cluster temperature distribution function (Colafrancesco et al. 1997).
Observations of very hot clusters can alleviate this problem to a
certain extent.  The cluster temperature function is predicted to have
an exponential cutoff beyond a critical temperature which depends on
the cosmological model, so the differences between the theoretical
predictions increase exponentially beyond the critical temperature.

Here we report on optical and X-ray observations of a
newly discovered cluster, 1E0657-56.  The measured temperature
$\rm{kT}=17.4 \pm2.5 \rm{keV}$ makes it a strong 
candidate for the hottest known cluster.  The interpretation of this temperature 
is complicated by evidence that the cluster is in the process of merging 
with a subcluster (see e.g. Schindler \&
Muller, 1993, Roettiger et al 1997) .  Since the subcluster has a mass on the 
order of one-tenth the mass of the primary cluster, we argue that the high 
temperature is a true indication of a large primary mass.  If this interpretation 
is correct, then the existence of such a cluster at the observed redshift
$z=0.296$ is very difficult to reconcile with current theoretical
models for a critical density ($\Omega = 1$)  universe.
     
This letter is organized in the following manner.
The optical and X-ray observations are presented and analyzed in
$\oint 2$, and the implications of the exceptionally high temperature of the
hot gas in 1E0657-56 are considered in $\oint 3$.

\section{OBSERVATIONS}
\subsection{Optical Observations}

The cluster 1E0657-56, was detected as an extended source on a scale of $3^{\prime}$
by the Imaging Proportional Counter (IPC) onboard the Einstein Observatory.  The source 
remained unidentified until a CCD image of this source was obtained with the Las
Campanas 2.5 meter telescope as part of a program to
search for possible failed clusters of galaxies, i.e., clusters
with hot gas but no galaxies (Tucker, Tananbaum \& Remillard 1995). The CCD
image revealed the presence of a rich cluster of galaxies and a luminous arc 
about 12 arcsec long.  Follow-up imaging and spectroscopic observations with the 
ESO NTT telescope in December 1993 confirmed the presence of the arc and 
yielded a cluster redshift of $z = 0.296$. The
corresponding luminosity distance is $D= 2.04~h_{50}^{-1}$~Gpc for $q_{0}= \Lambda = 0$,
where $h_{50}$ is the Hubble  constant in units of 50 km~s$^{-1}$~Mpc$^{-1}$. 

We use our NTT spectroscopic data to calculate the cluster velocity dispersion.
The method of  Danese et al (1980) is used 
to include the effects of measurement errors. From the 24 measured
redshifts we only include galaxies with velocities between 80,000~km~s$^{-1}$,
and 100,000~km~s$^{-1}$, and within
$R_{max} = 3$\ arcmin of the brightest cD galaxy in the 
cluster.  The resulting velocity dispersion is $\sigma_{v} =
1213^{+352}_{-191}$~km~s$^{-1}$, based on  13 galaxies. When we vary
$R_{max}$ between 1\ arcmin and 4\ arcmin,
we find $\sigma_{v}$ varies from 830 to 1515~km~s$^{-1}$, respectively, and 
the corresponding number of galaxies varies from 7 to 15. At the lower
limit, the number of galaxies is too small to yield useful
errors; at the upper limit, the errors are comparable to those
with $R_{max} =3$\ arcmin.

\subsection{ROSAT HRI Observations}

The ROSAT HRI observed 1E0657-56 for a total live
time of 57.7 ksec between 8 February 1995 and 27 June 1995.  A
contour image of the background subtracted HRI image is 
shown in Figure 1.  This figure
shows that the X-ray emission is extended and double-peaked,
indicating either a merger event or two spatially distinct 
groups.   The X-ray flux of the secondary peak, which is also 
an extended source, comprises only 4\% of the total flux.

The surface brightness profile of the primary cluster
was obtained by extracting the net counts in concentric annuli
about the primary X-ray peak after excluding the western quadrant
(which contains the secondary peak) and all point sources.  We 
then fit the observed X-ray surface brightness profile
to the standard hydrostatic-isothermal $\beta$ model,

$$\Sigma(R) = \Sigma_0(1+(R/a)^{2})^{-3\beta +1/2}$$

\noindent
with the addition of a flat background component.  
The best fit is obtained with a core radius of 
$a$ = 62$^{\prime\prime}\pm10^{\prime\prime}$ (corresponding to
$360 \pm50 h_{50}^{-1}$ kpc), $\beta = 0.62\pm0.07$,
and a central electron number density 
$n_e = 7.5 \times 10^{-3} h_{50}^{1/2}$~cm$^{-3}$. 
Integrating the best fit $\beta$ model gives a
gas mass within $1 h_{50}^{-1}$~Mpc of 
$2.0 \pm 0.3 \times 10^{14} h_{50}^{-5/2} M_{\odot}$.
The net flux and luminosity of the primary cluster is then determined by 
multiplying the flux
from the 3 quadrants centered on the primary peak by 4/3 to account for 
the flux underlying the secondary peak.  Using the gas temperature
derived from the ASCA data gives 
$f_{0.1-2.4~\rm{keV}} = 5.6 \pm 0.6 \times 10^{-12}$ erg cm$^{-2}$ s$^{-1}$
and $L_{bol} = 1.4 \pm 0.3 \times 10^{46} h_{50}^{-2}$ erg s$^{-1}$.

\subsection{ASCA Observations}
     
Standard screening criteria were applied to the SIS and GIS data to 
exclude data taken during Earth occultations, periods
of high particle flux, and the portions of ASCA orbit that pass
through the South Atlantic Anomaly.  Hot and flickering pixels
were removed with SISCLEAN and GISCLEAN. 
The resulting screened exposure times are: SIS0 (20.7 ksec); SIS1 (20.0 ksec);
GIS2 (21.7 ksec); and GIS3 (21.7 ksec).
Source spectra were extracted using a rectangular region of 
$4.2^{\prime}$ by $3.7^{\prime}$ for the SIS0 data, a 
rectangular region of $4.2^{\prime}$ by $3.0^{\prime}$ for the 
SIS1 data, and circular regions of radii $6.4^{\prime}$ and 
$7.1^{\prime}$ for the GIS2 and GIS3 data, respectively.
Background spectra were taken from available blank sky observations.

All four background subtracted spectra are then simultaneously fit
to a single-temperature Raymond thermal plasma model. The best
fit model ($\chi^{2}=692$ for 778 spectral bins) and residuals
are shown in Figure 2a.  The best fit model has a temperature of 
$\rm{kT} = 17.4\pm 2.5 \rm{keV}$ (90\% confidence level) and
an Fe abundance of $0.35\pm 0.15$ solar ($\rm{Fe/H} = 4.68 \times 10^{-5}$).
A red-shifted 6.9 keV line, corresponding
to the FeXXVI $K_{\alpha}$ line is also detected, but only at the
2.5 $\sigma$ level. Figure 2b shows a comparison of the SIS0 data 
from 1E0657-56 with the best fit single-temperature Raymond model 
for A2163 ($\rm{kT} = 12\pm 2$ \rm{keV};
Markevitch et al 1996). 1E0657-56 is clearly hotter than A2163.
The best fit temperature is also greater than 
that of MS1054-53, reported to have a temperature of
$14.7^{+4.6}_{-3.5}$ keV (Donahue et al 1997).  At $\rm{kT} = 17.4$ keV,  1E0657-56 is 
to our knowledge the hottest known cluster of galaxies.
For a temperature of 17 keV and the velocity dispersion
given above, we find $\beta_{spec}=\mu m_{p}\sigma_{v}^{2}/kT =
0.56$, which agrees well with the value of $\beta$ determined from the
fit to the HRIHRI  surface brightness profile.

An absorbed power-law model (photon index = 1.5$\pm$ 0.05) also provides 
an acceptable fit to the ASCA data, with a reduced chi-squared of 0.932.  
However the ROSAT HRI image shows that 1E0657-56 is extended in the 0.1 to
2.4 keV band, so a pure power law from an active galactic
nucleus (AGN) cannot explain the entire 0.5 to 10 keV ASCA spectrum. 
The FWHM of the ASCA/SIS point response function
is comparable to the extent of the cluster, so we cannot
determine if the hard emission outside of the PSPC bandpass is extended.
However, the centroids of the ASCA image in both the 1-4 keV and
4-10 keV bands are coincident with the centroid of the ROSAT image,
supporting the assumption that the ROSAT and ASCA
emission have the same origin, namely the hot cluster gas. 
Assuming the gas is isothermal gives a total 
gravitating mass within $1~h_{50}^{-1}$ Mpc of 
$1.1 \pm 0.2 \times 10^{15} h_{50}^{-1} M_{\odot}$, 
based on the equation of hydrostatic equilibrium, the best fit
$\beta$ model, and the ASCA determined gas temperature. This gives
a gas mass fraction of 
$0.18 ~h_{50}^{-3/2}$, consistent with other rich clusters
(David, Jones, \& Forman 1995; White \& Fabian 1995).

\section{ DISCUSSION}

The thermal energy content of the cluster gas within a radius
of $1 h_{50}^{-1}$~Mpc is approximately $2 \times 10^{64}$ ergs.  It
is difficult to account for this energy content by any process
other than gravitational collapse.  Assuming a thermal energy
input of 10$^{51}$ ergs per supernova, $1.5\times 10^{13}$
supernovae would be required to heat the gas.  Assuming an equal
mixture of Type Ia supernovae (Fe yield of $0.6 M_{\odot}$) and
Type II  supernovae (Fe yield of $0.08 M_{\odot}$),
the resulting iron abundance 
would be more than ten times the solar abundance and
thirty times greater than implied by the ASCA data.  A quasar can
generate $\sim 10^{46}$ ergs  s$^{-1}$ of energy for
$\sim 10^{9}$ years; ten such quasars would be required to explain
the thermal energy content of 1E0657-56, but then the space density of
quasars in the cluster would have to be $\sim$100 times greater
than the cosmic average.  

The cluster is observed to be a radio source; Griffith and Wright
(1993) detected a 1.7mJy radio source at 4850 MHz.
Inverse Compton scattering of the microwave background off the
relativistic electrons in the radio source could produce an
extended, hard X-ray source.  The intensity of this source in the 
0.1-10 keV band is of the order of 

\begin{eqnarray}
L_{C}(0.1-10~keV) & \sim &  L_{syn}(8\pi
w_{bb}/B^{2})(\nu_{x}/\nu_{bb}\gamma^{2})  \\
                  & \sim & 6 \times 10^{39}(10^{-6}\rm{gauss}/B)~{\rm erg~s^{-1}} 
\end{eqnarray}
 
\noindent
However, since the equipartition magnetic field is $B_{eq} \sim 10^{-6}$
gauss, the inverse Compton contribution to the X-ray intensity is negligible.

Numerical simulations of merging clusters (e.g. Roettiger, et
al. 1997, Evrard et al. 1996, Schindler \& Muller, 1993) show that
temperatures of order 20 keV can be achieved in mergers, a conclusion
confirmed by the ASCA observations of Abell 754 (Henriksen \&
Markevitch 1996). However, the simulations and the observations both
indicate that for mergers with high cluster/subcluster mass ratio, the
high temperature regions are usually localized, and do not affect
significantly the average cluster temperature.  In A754, the average
temperature is $\sim 9$~keV, even though a temperature in the range $> 12$~keV
is found in one region (Henriksen \& Markevitch 1996).  This is in
agreement with the numerical simulations of Roettiger et al.( 1997)
for a cluster/subcluster mass ratio of 8:1.  Their calculations show
little change of the emission weighted temperature, except for a peak
at the time of the merger which is confined to the central 300 kpc for
roughly a few hundred million years.  For 1E0657-56, we can estimate
the mass ratio from the HRI observation, which indicates that 4\% of
the flux comes from the secondary peak which occupies ~10\% of the
volume, $V$.  Since the luminosity $L \propto M^{2}/V$, the mass ratio
is $\sim$ 16:1, so the effect of the merger on our emission weighted
temperature should not be large for this particular cluster.

One of the strongest correlations between the X-ray properties
of clusters of galaxies is that between the gas temperature and 
X-ray luminosity (David et al. 1993; Mushotzky \& Scharf 1997). 
Figure 3 shows that, within the errors, the temperature of 
1E0657-56 is consistent with this relation, providing
further support that the high temperature reflects
a global property of the cluster, and is not due to a small
scale temperature enhancement.

In summary,  the numerical simulations do illustrate the importance 
of having spatial as well as spectral information before using 
cluster temperatures as an indicator of virial masses, but in the
present case we conclude that the unusually high temperature of 
1E0657 is due to an unusually deep potential well and hence a large virial mass.

Every cluster that is discovered with a temperature greater than about
15 keV places an additional straw on the "camel's back" of
cosmological theories for the formation of clusters of galaxies that
require $\Omega$ = 1. The standard normalization for primordial
density fluctuations is derived from the number density of 5-8~keV
clusters (Henry \& Arnaud 1991, Eke et al 1996, Pen 1996, Viana \&
Liddle 1996).  Differences between cosmological models are strongly
amplified at temperatures of 17 keV.  For example, the expected number
of clusters with a temperature greater than 17.4 keV, in a volume
bounded by a redshift $z= 0.3$ is $N(T>17.4,z<0.3) = 0.3$ for $\Omega
= 0.37$ and $N(T>17.4,z<0.3) = 4 \times 10^{-6}$ for $\Omega = 1.00$
for a CDM spectrum normalized to fit the cluster number density at 5.5
keV (U. Pen, private communication). The discovery of just one high
temperature cluster thus effectively rules out models with a cold dark
matter (CDM) type of power spectrum of fluctuations and $\Omega=1$.
Other authors (Donahue et al. 1997, Luppino \& Gioia 1995, Ostriker
1993, and references cited therein) have noted that standard CDM models
with $\Omega=1$ cannot explain the COBE measurements of CMBR
temperature anisotropies (Smoot et al 1992), as well as the abundances
of clusters and galaxies.  
 
In the future we can anticipate that the discovery and
study of very hot gas in high redshift clusters of galaxies with
ASCA and future observatories such as AXAF and XMM will provide
severe tests for cosmological models for the formation of
large-scale structure in the universe.  

\acknowledgments

We are grateful for the assistance of K. Arnaud, P.  Gorenstein,
U. Pen, and R. Mushotzky. This research was supported by NASA Contract
NAS8-39073 and the Smithsonian Institution. We benefited from the use
of the NASA Extragalactic Data Base to locate the radio source
identified with 1E0657-58.

\clearpage

\figcaption{Contour map of the ROSAT HRI image for  1E0657-56.
The contour map is generated from a background subtracted, wavelet
smoothed image using the technique described in
Vikhlinin, Forman, and Jones (1994).
The peak contour level is drawn at a
surface brightness of 5.9 $\times$ 10$^{-2}$~cts~arcmin$^{-2}$~sec$^{-1}$.
The remaining contour levels are spaced at factors of two decrease
in surface brightness. The lowest contour is $4 \sigma$ above background.
\label{fig1}}

\figcaption{a) Best fit Raymond thermal plasma model to the ASCA data.
Filled squares (SIS), Open squares (GIS). b) Comparison of the
SIS0 spectrum of 1E0657-56 with the best fit spectrum for A2163
(solid line). The 1E0657-56 data points above 6 keV diverge noticeably
from the A2163 fit, confirming that 1E0657-56 is hotter than A2163.\label{fig2}}

\figcaption{The luminosity-temperature plot for clusters of
galaxies, adapted from David et al.1993 and Mushotzky \& Scharf 1997, showing the location of
A2163, MS1054-53 and 1E 0657-56. \label{fig3}}

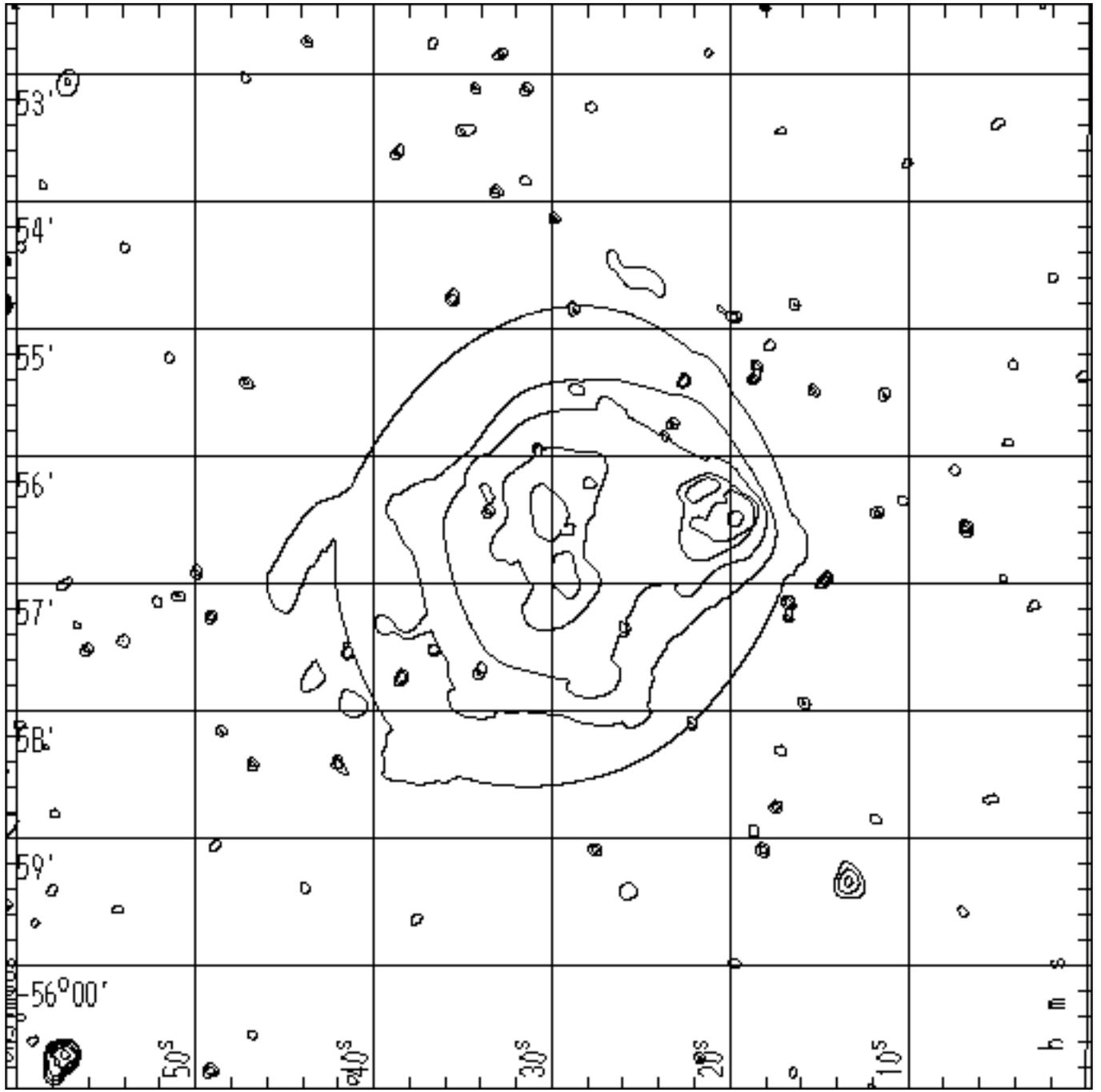
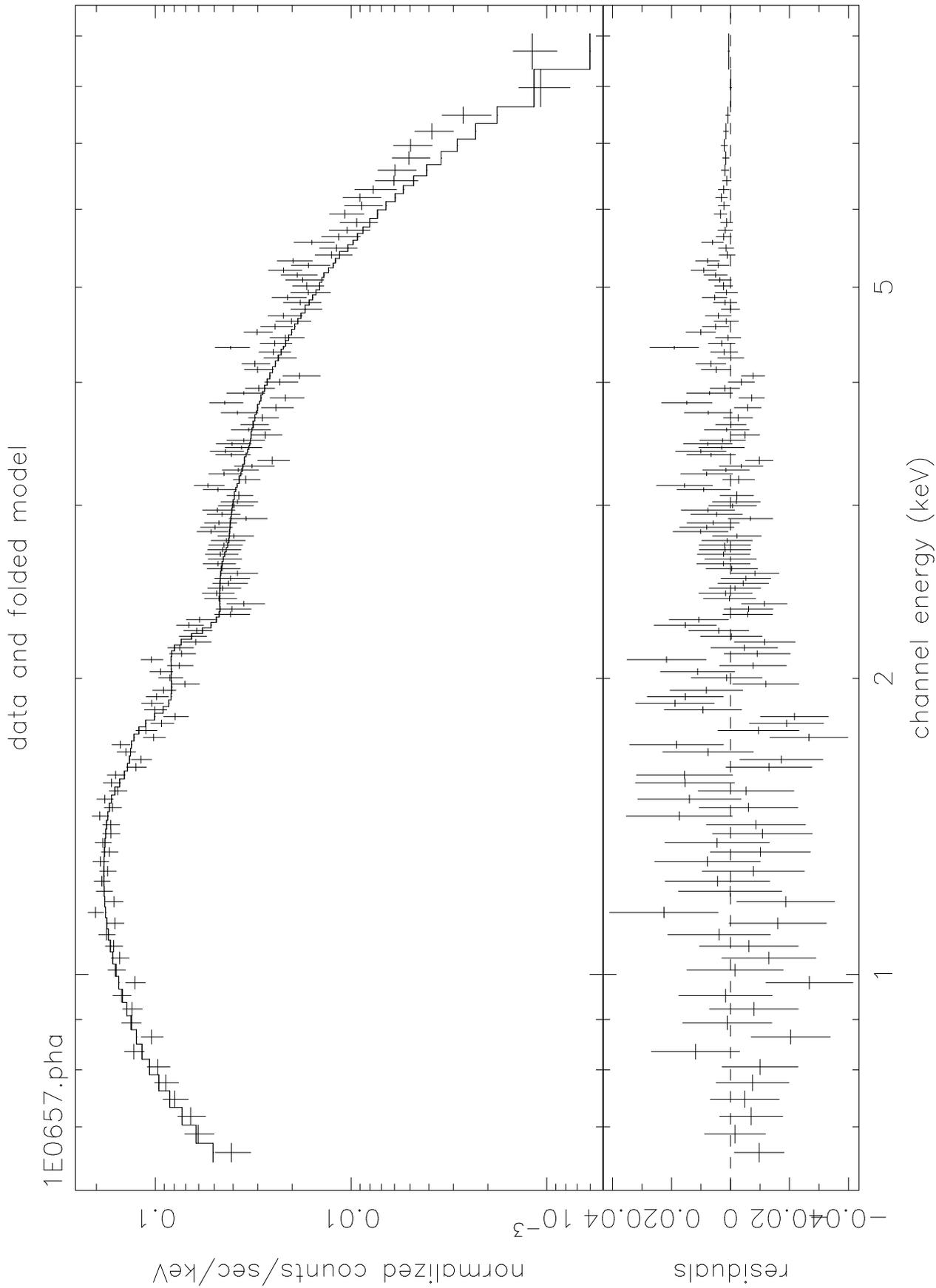
\end{document}